\documentclass[aps,pre,twocolumn,floatfix,superscriptaddress]{revtex4-1}
\usepackage[pdftex]{hyperref,graphicx}	
\usepackage{dcolumn}   	\usepackage[pdftex]{hyperref,graphicx}	
\usepackage{bm}
\usepackage{natbib}
\usepackage{array}
\usepackage{inputenc}
\usepackage[active]{srcltx}

\begin{document}
\title{Analytical Solution of the Proprio-Graviceptive equation for shoot gravitropism of plants.}
\author{Renaud Bastien}
\email{renaud.bastien@clermont.inra.fr}
\affiliation{Institut Jean-Pierre Bourgin, UMR1318 INRA-AgroParisTech, 78026 Versailles, France}
\affiliation{INRA, UMR 547 PIAF, F-63100 Clermont Fd Cedex 01}
\affiliation{Clermont Universit\'e, Universit\'e Blaise Pascal, UMR 547 PIAF, BP 10448, F-63000 Clermont-Ferrand}
\affiliation{Mati\`ere et Syst\`emes Complexes, Universit\'e Paris-Diderot, 10 rue Alice Domont et L\'eonie Duquet, 75025 Paris Cedex 13, France}
\author{Bruno Moulia}
\affiliation{INRA, UMR 547 PIAF, F-63100 Clermont Fd Cedex 01}
\affiliation{Clermont Universit\'e, Universit\'e Blaise Pascal, UMR 547 PIAF, BP 10448, F-63000 Clermont-Ferrand}
\author{St\'ephane Douady}
\affiliation{Mati\`ere et Syst\`emes Complexes, Universit\'e Paris-Diderot, 10 rue Alice Domont et L\'eonie Duquet, 75025 Paris Cedex 13, France}
\author{Tomas Bohr}
\affiliation{Center of Fluid Dynamics and Department of Physics, Technical University of Denmark, 2800 Kgs.~Lyngby, Denmark}



\begin{abstract}
We derive  the analytical solutions to the second order generalised gravi-proprioceptive equation given in our recent paper \cite{Bastien2012}.  These equations show how plants adjust to the surrounding gravitation field and highlight the fact that the plant must be able to not only sense its local posture with respect to the gravitational field, but also to sense its own local curvature. In  \cite{Bastien2012} we obtained explicit analytical solutions of these equations in terms of (sums of) Bessel functions, and in the present paper we derive these solutions.
\end{abstract}
\maketitle
Gravitropism is a slow movement by which plants reorient their  growth  in response to gravity. In a recent paper \cite{Bastien2012}, we have studied  the gravitropic response of a broad range of plants by placing them horizontally in a dark room and monitoring how they curve upwards in response to gravity. It was shown that the minimal dynamical model accounting for these observations is not the usual gravisensing law, where the plant simply senses the local tilt angle, but what we have called a {\em gravi-proprioceptive} law, where one takes into account that the plant can, in addition, sense its own local curvature. In that paper, exact solutions for the linearized versions of both the graviceptive and the gravi-proprioceptive law were given and compared to experiment. We here give the derivation of these solutions

The plant is described as a slender rod with arc length $s$ going from $s=0$ at the base to $s=L$ at the tip. The local tilt angle $A(s,t)$ describes the orientation with respect to the vertical gravitational field, such that $A=0$ is vertical.
The generalized (gravi-proprioceptive) law of gravitropism describes how the local curvature $C(s,t))$ changes in response to the local tilt angle {\em and} the local curvature itself as
\begin{equation}
\label{eqnbase}
\frac{\partial C(s,t)}{\partial t}=-\beta A(s,t)-\gamma C(s,t),
\end{equation}
where the curvature and the tilt angle are related by
\begin{equation}
C(s,t)=\frac{\partial A(s,t)}{\partial s},
\end{equation}
and where the coefficient $\beta$ controls the graviceptive term (sensing of tilt angle) and the coefficient $\gamma$ controls the gravi-proprioceptive term (sensing the curvature).

The subject of the present paper is to solve this equation with the experimentally relevant boundary and initial condition. Thus we assume that the plant is initially straight with an angle, say $A_i$ from the vertical. Since (\ref{eqnbase}) is {\em linear} in $A$, we can, without loss of generality, take $A_i=1$. The solution with a different $A_i$ will simply be obtained by multiplying our solution by $A_i$.Thus we assume the initial conditions
\begin{equation}
\label{bc1}
A(s,t=0)=1
\end{equation}
and 
\begin{equation}
\label{bc2}
C(s,t=0)=\frac{\partial A(s,t=0)}{\partial s}= 0
\end{equation}
for all $0<s<L$. Further we assume that the plant is {\em clamped} in the sense that tilt angle at the base remains equal to $A_i =1$ for all future times. Thus the boundary condition is
\begin{equation}
\label{bc3}
A(s=0,t)=1
\end{equation}
for all $ t \ge 0$

\section{The Graviceptive model \label{modA}}
We first consider the purely graviceptive model, i.e., (\ref{eqnbase}) with $\gamma=0$:
\begin{eqnarray}
\label{base0}
\frac{\partial C(s,t)}{\partial t}=-\beta A(s,t)
\end{eqnarray}
To solve this equation, we first set $\beta=1$ with out loss of generality simply by using the scaled time $\beta t$. Now, using that 
\begin{equation}
C(s,t) = \frac{\partial A(s,t)}{\partial s}
\end{equation}
we can write (\ref{base0}) as
\begin{eqnarray}
\label{base0A}
\frac{\partial^2 A(s,t)}{\partial s \partial t} = \partial^2_{st} A=- A
\end{eqnarray}
where we use the short hand notation $\partial/ \partial x = \partial_x$. Note that $C$ satisfies the same equation.

\subsection{Solution by separation}
We shall now try to find separated solutions of (\ref{base0A}). From our numerical work it seems that the solutions depend on $\sqrt{s}$ and $\sqrt{t}$ and 
since both $s$ and $t$ are assumed positive, we can introduce the new independent variables  $(\xi, \eta)$ as
\begin{eqnarray}
\xi = (s\, t)^{1/2} \,\,\,\,\,\,\,{\rm and}\,\,\,\,\,\,\, \eta = \left( \frac{t}{s} \right)^{1/2}
\end{eqnarray}
or 
\begin{eqnarray}
s =  \frac{\xi}{\eta}  \,\,\,\,\,\,\,{\rm and}\,\,\,\,\,\,\,t =\xi \eta
\end{eqnarray}
and obtain
\begin{eqnarray}
\partial_s &=& (\partial_s \xi) \,  \partial_ {\xi} +     (\partial_s \eta) \, \partial_ {\eta} = \frac12 \eta \, \partial_{\xi} - \frac12 \eta^2 \xi^{-1}\, \partial_{\eta}\\
\partial_t &=&  (\partial_t \xi) \,  \partial_ {\xi}+     (\partial_t \eta) \,  \partial_ {\eta} = \frac12 \eta^{-1} \, \partial_{\xi} + \frac12  \xi^{-1}\, \partial_{\eta}
\end{eqnarray}
Using this, we find
\begin{equation}
 \partial^2_{st} = \frac{1}{4 \xi^2} \left(  \xi^2 \partial^2_{\xi \xi} + \xi \partial_{\xi} - \eta \partial_{\eta} + \eta^2  \partial^2_{\eta \eta}     \right)
\end{equation}
Note that the $\partial^2_{\xi \eta}$ cancel and that the invariance of (\ref{base0}) on exchanging $s$ and $t$ implies an invariance when $\eta \rightarrow \eta^{-1}$.
In these variables, (\ref{base0A}) for an arbitrary function $\Psi$ can be written
\begin{equation}
 \left( \xi^2 \partial^2_{\xi \xi} + \xi \partial_{\xi} - \eta \partial_{\eta} + \eta^2  \partial^2_{\eta \eta}    + 4 \xi^2 \right) \Psi(\xi,\eta)=0
\end{equation}
If we now assume that the solution can be written in the separated form
\begin{equation}
 \Psi(\xi, \eta)  = f(\eta) g(\xi)
\end{equation}
we get the two eigenvalue equations
\begin{eqnarray}
\label{ev1}
 \xi^2 g''(\xi) + \xi \, g'(\xi) + 4 \xi^2 g(\xi)&=& \lambda \,  g(\xi)\\
 \label{ev2}
\eta^2 f''(\eta) - \eta f'(\eta) &=&- \lambda \, f(\eta)
\end{eqnarray}
Substituting $y = \log \eta$ gives (\ref{ev2}) the form
\begin{equation}
f''(y) - 2 f'(y) + \lambda \, f(y) = 0
\end{equation}
with solutions
\begin{equation}
f(y) = e^{p \, y} = \eta^p
\end{equation}
with
\begin{equation}
p^2 - 2 p + \lambda = 0
\end{equation}
or
\begin{equation}
p=\left( 1 \pm \sqrt{1- \lambda}) \right)
\end{equation}
and then  (\ref{ev1}) becomes
\begin{equation}
 \xi^2 g''(\xi) + \xi \, g'(\xi) + \left (4 \xi^2 - p(2-p)\right)g(\xi)  = 0
\end{equation}
If we finally substitute $x = 2 \xi$,  this becomes
\begin{equation}
\label{Bessel}
x^2 g''(x) + x \, g'(x) + \left (x^2 - p(2-p)\right)g(x)  = 0
\end{equation}
which is recognised as Bessel's equation of order $n$, where $n^2=p(2-p)$. For $n$ to be a real number we must restrict $p$ to lie in the interval
$p \in [0,2]$.
In this interval we can write the full solution as
\begin{equation}
\label{psi}
\Psi_n(s,t) = K  \left( \frac{t}{s} \right)^{p/2} J_n (2 \sqrt{s\, t}  )
\end{equation}
If $n$ is to be an integer we must choose $p=0$, 1 or 2 giving $n=0$, 1 and 0, respectively.
To find a solution for $A$, we must satisfy the boundary condition $A (s \rightarrow 0,t) = A (s ,t\rightarrow 0)=1$ and since
\begin{equation}
 \left( \frac{t}{s} \right)^{p/2}  J_n (2 \sqrt{st})  \rightarrow s^{(n-p)/2}  t^{(n+p)/2} 
\end{equation}
for small $s \, t$, we must choose $n=p=0$, i.e., 
\begin{equation}
A(\xi,\eta) = J_0 (2 \xi)
\end{equation}
or, returning to the variables  $s$ and $t$ and re-inserting $\beta$,
\begin{equation}
\label{sol01}
A(s,t) = J_0 (2 \sqrt{\beta s\, t} )
\end{equation}
 We can also find the curvature
\begin{equation}
\label{sol02}
C(s,t) = \partial_s A(s,t)=   \sqrt{\frac{\beta t}{s}}J_{1}\left(2\sqrt{\beta t s}\right)
\end{equation}
which, since $C$ also satisfies (\ref{base0A}), also has the form (\ref{psi}). For small $s$, the $s$-dependence cancels since $J_1(x) \approx x/2$ for small $x$, and we get
\begin{equation}
C(s,t) \rightarrow  \beta t  
\end{equation}
for small $s$ or $t$ and thus $C(s,t\rightarrow 0) =0$ as we demanded.

\subsection{Solution by Laplace transformation}
We could have solved (\ref{base0A}) more directly  Laplace transforming is  (in $s$), which gives
\begin{eqnarray}
p \partial_t \hat{A} - \partial_t A(s=0,t) = -\beta \hat{A} ,
\end{eqnarray}
where
\begin{equation}
\hat{A}(p,t)=\int_0^{\infty} A(s,t) e^{-ps} ds
\end{equation}
is the Laplace transform of $A(s,t)$. From the boundary condition (\ref{bc3}) we get $ \partial_t A(s=0,t)=0$ and 
\begin{equation}
 \partial_t\hat{A}  = -\frac{\beta}{p} \hat{A}, 
\end{equation}
which has to be solved with the initial condition $A(s,t=0)=1 \Rightarrow \hat{A(p,t=0)}=1/p$
with the solution
\begin{eqnarray}
\nonumber
\hat{A} &=&\frac{1}{p} e^{-\frac{\beta t}{p} }
\end{eqnarray}
We see that $\hat{A} (p,t)$ is analytic in the complex $p$-plane except at the isolated (essential) singularity $p=0$.
The inverse Laplace transform $A(s,t)$ can simply be found as the residue (see e.g. \cite{Marsden1999} Theorem 8.2.1)
\begin{equation}
\label{J0}
A(s,t)= \frac{1}{2 \pi i}\oint_{C(0,\epsilon)} \frac{e^{sp -\frac{\beta t}{p}}}{p}dp = J_0(2\sqrt{\beta s t})
\end{equation}
where the last equality is a special case of the well-known identity derived e.g. in \cite{Watson1944}:
\begin{equation}
\label{Watson}
F_n(s,t)= \frac{1}{2 \pi i}\oint_{\Omega(0,\epsilon)} \frac{e^{p s-\frac{t}{p}}}{p^{n+1}}  dp = \left(\frac{s}{t}\right)^{n/2} J_n(2 \sqrt{st}\,)
\end{equation} 
valid for any $n>-1$. 

\section{The Proprio-graviceptive model\label{modC}}

We now consider the generalised equation of gravitropism 
\begin{eqnarray}
\label{eqnbase31}
 \partial_t C(s,t)=-\beta A(s,t)-\gamma C(s,t)
\end{eqnarray}
and with scaled time $T = \gamma t$ and $S=(\beta/\gamma) s$ we obtain
\begin{eqnarray}
\partial^2_{ST} A=- A-  \partial_S A
\end{eqnarray}
In the following we drop the capitals  and our equation becomes
\begin{eqnarray}
\label{eqnbasepsi0}
\partial^2_{st } A=- A-  \partial_s A
\end{eqnarray}
again with the initial conditions (\ref{bc1}) and (\ref{bc2}) and the clamped boundary condition (\ref{bc3}).

Note that (\ref{eqnbasepsi0}) has that stationary (asymptotic) solution 
\begin{equation}
\label{stat}
A_0(s) = e^{-s}
\end{equation}
satisfying the boundary conditions~\ref{bc1} and~\ref{bc2}. One might try the substitution $A(s,t) = B(s,t) e^{-t}$, since the equation for  $B$ would simply be (\ref{base0}). However, the boundary condition (\ref{bc3}) would be $B(s=0,t) = e^t$ which is time dependent and complicates matters.

The Laplace transform is
\begin{equation}
\hat{A}(p,t)=\int_0^{\infty} A(s,t) e^{-ps} ds
\end{equation}
and the Laplace transform of~\ref{eqnbasepsi0} (with $\gamma = \beta = 1$) gives
\begin{eqnarray}
p \partial_t\hat{A} -\partial_t A(s=0,t) = -\hat{A} - (p \hat{A} - A(s=0,t))
\end{eqnarray}

From the boundary condition, we know that $A(s=0,t)$ is constant in time and thus $\partial_t A(s=0,t)=0$. Also
$A(s=0,t) = 1$ so we obtain
\begin{equation}
p \partial_t \hat{A} = -\hat{A} - p \hat{A} +  1
\end{equation}
or
\begin{equation}
\partial_t \hat{A} +\frac{1+p}{p}\hat{A} = \frac{1}{p} 
\end{equation}
with the solution
\begin{eqnarray}
\nonumber
\hat{A} &=& \frac{1}{1+p} + \frac{1}{p(1+p)} e^{-\frac{1+p}{p} t} \\
\nonumber
&=&  \frac{1}{1+p} + \left( \frac{1}{p} -  \frac{1}{1+p}\right)e^{-t} e^{-\frac{t}{p}}\\
&=&  \frac{1}{p} e^{-t} e^{-\frac{t}{p}} +  \frac{1}{1+p} \left(  1-e^{-t} e^{-\frac{t}{p}}        \right) ,
\end{eqnarray}
where $\hat{A} (p, t=0) = 1/p$ in accordance with the boundary condition. It correctly approaches the stationary state 
since
\begin{equation}
\hat{A}(p,t) \rightarrow   \frac{1}{1+p} =\hat{A_0} 
\end{equation}
for $t \to \infty$ when $p>0$ and  $\hat{A} (p,t)$ is analytic in the complex $p$-plane except at the isolated singularities $p=0$ and $p=-1$. 

Thus the inverse Laplace transform $A(s,t)$ is simply the sum of the residues \cite{Marsden1999} of the function 
\begin{eqnarray}
\nonumber
A(s,t) = \sum_{\rm{singularities} \, p^*}&&\frac{1}{2 \pi i}\oint_{\Omega(p^*,\epsilon)} e^{p s}\left(\frac{1}{p} e^{-t} e^{-\frac{t}{p}} \right.\\
&+& \left. \frac{1}{1+p} \left(  1-e^{-t} e^{-\frac{t}{p}}        \right)\right)
\end{eqnarray}
 at each of its singularities in the complex $p$-plane. The pole at $p=-1$ is a simple pole, but the residue is zero since the numerator $1-e^{-t} e^{-\frac{t}{p}}$ is zero for $p=-1$.  We therefore only need  the residue at $p=0$, which is an essential singularity. Thus we have to evaluate the contour integral of $e^{p s} \hat{A}(p,t)$ over a closed curve $\Omega(0,\epsilon)$ encircling the origin. The contribution of the term $e^{ps}/(1+p)$ vanishes, since it has no pole at $p=0$, and we can write
\begin{equation}
A(s,t)= A_1(s,t) + A_2(s,t) ,
\end{equation} 
where
\begin{equation}
\label{J0e}
A_1(s,t)= \frac{e^{-t}}{2 \pi i}\oint_{\Omega(0,\epsilon)} \frac{e^{sp -\frac{t}{p}}}{p}dp
\end{equation} 
and 
\begin{equation}
A_2(s,t)= -\frac{e^{-t}}{2 \pi i}\oint_{\Omega(0,\epsilon)}  \frac{e^{sp -\frac{t}{p}}}{1+p} dp,
\end{equation} 
where first term is given by (\ref{J0})
\begin{equation}
\label{A_1}
A_1(s,t)= e^{-t}J_0(2 \sqrt{st}\,).
\end{equation} 

Near the origin (more precisely, as long as $\mid p \mid < 1$) we can expand
\begin{equation}
\frac{1}{1+p} = 1 - p + p^2  \ldots = \sum_{n=0}^{\infty} (-p)^n
\end{equation}
so that
\begin{eqnarray}
\nonumber
A_2(s,t) = -\frac{e^{-t}}{2 \pi i}\sum_{n=0}^{\infty}&& \oint_{\Omega(0,\epsilon)}  e^{sp -\frac{t}{p}} (-p)^n dp \\
&=& e^{-t}\sum_{n=0}^{\infty} T_n(s,t),
\end{eqnarray} 
where
\begin{equation}
T_n(s,t)= -\frac{1}{2 \pi i} \oint_{\Omega(0,\epsilon)}  e^{s p -\frac{t}{p}} (-p)^n dp.
\end{equation} 
We define $z=-1/p$ and thus $dp = dz/z^2$, whereby 
\begin{eqnarray}
\nonumber
T_n(s,t) &=& \frac{1}{2 \pi i} \oint_{\Omega(0,\epsilon)}  e^{t z  -\frac{s}{z}} z^{-(n+2)} dz \\
&=&  \left(\frac{t}{s}\right)^{\frac{n+1}{2}} J_{n+1}(2 \sqrt{st}\,)
\end{eqnarray} 
for any $n>0$. Note the sign change since the contour $\Omega$ is now traversed in the {\em clockwise} direction. Thus we can write
\begin{eqnarray}
\nonumber
A_2(s,t)&=& e^{-t}\sum_{m=0}^{\infty}  \left(\frac{t}{s}\right)^{\frac{m+1}{2}} J_{m+1}(2 \sqrt{st}\,) \\
&=&  e^{-t}\sum_{n=1}^{\infty}  \left(\frac{t}{s}\right)^{\frac{n}{2}} J_{n}(2 \sqrt{st}\,) 
\end{eqnarray}
and the full solution is 
\begin{eqnarray}
\nonumber
A(s,t)&=& e^{-t}\sum_{m=0}^{\infty}  \left(\frac{t}{s}\right)^{\frac{n}{2}} J_{n}(2 \sqrt{st}\,) \\
&=&e^{-s}-\sum_{n=1}^{\infty}(-1)^n\left(\frac{s}{t}\right)^{n/2} J_{n}(2 \sqrt{st}\,)
\end{eqnarray}
which, going back to the original variables using  $t \to \gamma t$ and $s \to (\beta/\gamma) s$ gives
\begin{eqnarray}
\nonumber
&&A(s,t)= e^{-\gamma t}\sum_{m=0}^{\infty}  \left(\frac{\gamma^2 t}{\beta s}\right)^{\frac{n}{2}} J_{n}(2 \sqrt{\beta st}\,) \\
\label{final}
&=&e^{-\beta s/\gamma}-\sum_{n=1}^{\infty}(-1)^n\left(\frac{\beta s}{\gamma^2 t}\right)^{n/2} J_{n}(2 \sqrt{\beta st}\,)
\end{eqnarray}

If the initial angle is $A_i$, we get the same solution, but with the factor $A_i$ multiplying the solution (\ref{final}).
\bibliographystyle{plain}

\begin{thebibliography}{2}


\bibitem{Bastien2012}
R. Bastien, T. Bohr, B. Moulia and S. Douady: {\em A unifying model of shoot gravitropism reveals proprioception as a central feature of posture control in plants. } Submitted to PNAS  (2012)

\bibitem{Marsden1999} J. E. Marsden and M. J. Hoffman: {\em Basic complex analysis}. 3rd edition, Freeman (1999).

\bibitem{Watson1944}
G.N. Watson: {\em A Treatise on the Theory of the Bessel Functions.} 2nd edition,  Cambridge University Press (1944).


\end{thebibliography}

\end{document}